# A New Model for Calculating the Ground and Excited States Masses Spectra of Doubly Heavy Ξ Baryons

**NEDA MOHAJERY[1], NASRIN SALEHI[1*], HASSAN HASSANABADI[2]**

[1] *Department of Basic Sciences, Shahrood Branch, Islamic Azad University, Shahrood, Iran*

[2] *Physics Department, Shahrood University of Technology, Shahrood, Iran.*

*\*salehi@shahroodut.ac.ir*

In this study, since the doubly heavy baryons masses are experimentally unknown (except $\Xi_{cc}^+$ and $\Xi_{cc}^{++}$), we present the ground state masses and the positive and negative parity excited state masses of doubly heavy Ξ baryons. For this purpose, we have solved the six-dimensional hyperradial Schrödinger equation analytically for three particles under the hypercentral potential by using the ansatz approach. In this paper the hypercentral potential is regarded as a combination of the color Coulomb plus linear confining term and the six-dimensional harmonic oscillator potential. We also added the first order correction and the spin-dependent part contains three types of interaction terms (the spin-spin term, spin-orbit term and tensor term) to the hypercentral potential. Our obtained masses for the radial excited states and orbital excited states of $\Xi_{ccd}$, $\Xi_{ccu}$, $\Xi_{bbd}$, $\Xi_{bbu}$, $\Xi_{bcd}$ and $\Xi_{bcu}$ systems are compared with other theoretical reports, which could be a beneficial tool for the interpretation of experimentally unknown doubly heavy baryons spectrum.

## 1. Introduction

The doubly heavy baryons have two heavy quarks (c and b) with a light quark (d or u or s). The doubly heavy Ξ baryons family have up or down quarks but Ω family has a light strange quark and their masses spectra have been predicted in the quark model [1]. The SELEX collaboration announced only the experimental mass for the ground state of $\Xi_{cc}^+$ baryon and LHCb has determined the ground state of $\Xi_{cc}^{++}$ baryon mass while no triply heavy baryons have been observed yet [2]. Recently experiments and theoretical outcomes have been used in studying the heavy baryons. A lot of new experimental results have been reported by various experimental facilities like CLEO, Belle, BaBar, LHCb, etc [3, 4]. on ground states and many new excited states of heavy flavor baryons. Bottom baryons are investigated at LHC and Lattice-QCD whereas charm baryons announced at the B-factories [5, 6]. On the other hand the theoretical works are providing new results for doubly heavy baryons like the Hamiltonian model [7], relativistic quark model [8], the chiral unitary model [9], QCD sum rule [10, 11] and many more. Single- and double- heavy baryons in the constituent quark model were studied by Yoshida et al. They used a model in which there were two exceptions, a color-Coulomb term depending on quark masses and an antisymmetric L.S force. They studied on the low-lying negative-parity states and structures within the framework of a constituent quark model [7]. In Ref. [12], the authors, calculated the masses of baryons with the quadratic mass relations for ground and orbitally excited states. Wei et al estimated the masses of singly, doubly, and triply bottom baryons in Ref. [13]. Then studied on the linear mass relations and quadratic mass relations.

The light flavor dependence of the singly and doubly charmed states investigated by Rubio et al. They focused on searching the masses of charmed baryons with positive and negative parity [5]. In Ref. [14], the authors using lattice QCD for baryons containing one, two, or three heavy quarks. They applied nonrelativistic QCD for the bottom quarks and relativistic heavy-quark action for the charm quarks. Padmanath et al, determined the ground and excited state spectra of doubly charmed baryons from lattice QCD with dynamical quark fields [15]. The mass of the heavy baryons with two heavy b or c quarks for spin $\frac{1}{2}$ in the framework of QCD sum rules estimated by Aliev et al. They use the most general form of the interpolating current in its symmetric and anti-symmetric forms with respect to the exchange of heavy quarks, to calculate the two point correlation functions describing the baryons under consideration [16]. The authors calculated the masses and residues of the spin $\frac{3}{2}$ doubly heavy baryons within the QCD sum rules method In Ref. [17]. Eakins et al were ignored all spin-dependent interactions and assume a flavor independent potential, worked in the limit where the two heavy quarks are massive enough that their motion can be treated as essentially non-relativistic and QCD interactions can be well-described by an adiabatic potential [18]. The three-quark problem solved by Valcarce et al. by means of the Faddeev method in momentum space [19].

The masses of the ground and excited states of the doubly heavy baryons calculated by Ebert et al, baryons on the basis of the quark-diquark approximation in the framework of the relativistic quark model [20]. In Ref. [21], the authors, in the model with the quark-diquark factorization of wave functions estimated the spectroscopic characteristics of baryons containing two heavy quarks. Albertus et al, used five different quark-quark potentials that include a confining term plus Coulomb and hyperfine terms coming from one–gluon exchange. They solved the three–body problem by means of a variational ansatz made possible by heavy quark spin symmetry constraints [22].

In this study, we have used the hypercentral constituent quark model (hCQM) with Coulombic-like term plus a linear confining term and the harmonic oscillator potential [23]. We also added the first order correction and the spin-dependent part to the potential and calculation has been performed by solving six dimensional hyperradial Schrödinger equations by using the ansatz method. We have obtained the mass spectra of radial excited states up to 5S and orbital excited states for 1P-5P, 1D-4D and 1F-2F states.

This paper is organized as follows: we briefly remind the hypercentral constituent quark model and introduce the interaction potentials between three quarks in doubly heavy baryons in section 2. In sect. 3 we present the exact analytical solution of the hyperradial Schrödinger equation for our proposed potential. In sect. 4, our masses spectra results for ground, radial and orbital excited states of baryon family with six members are given and compare with other predictions. We present the conclusions in section 5.

## 2. Theoretical Framework: The HCQM Model and Hypercentral Potential

The hypercentral model has been applied to solve bound states and scattering problems in many various fields of physics. In this model, we consider baryons as three-body systems of constituent quarks. In the center of mass frame, the internal quark motion is described by the Jacobi coordinates ($\rho$ and $\lambda$) [24] and the respective reduced masses are given by

$$m_\rho = \frac{2m_1 m_2}{m_1 + m_2}, \qquad m_\lambda = \frac{2m_3(m_1^2 + m_2^2 + m_1 m_2)}{(m_1 + m_2)(m_1 + m_2 + m_3)} \tag{1}$$

Here $m_1$, $m_2$ and $m_3$ are the current quark masses. In order to describe three-quark dynamics, we define hyper radius $x = \sqrt{\rho^2 + \lambda^2}$ and hyper angle $\xi = \arctan\left(\frac{\rho}{\lambda}\right)$ [25]. In present work, the confining three-body potential is regarded as a combination of three hypercentral interacting potentials. First, the six-dimensional hyper-coulomb potential, $V_{hyc}(x) = \frac{\tau}{x}$, which is attractive for small separations [26, 27], while at large separations a hyper-linear term, $V_{con} = \beta x$, gives rise to quark confinement [28] where $\beta$ Corresponds to the string tension of the confinement [29]. Third, the six-dimension harmonic oscillator potential, $V_{h.o.} = px^2$, which has a two-body character, and turns out to be exactly hypercentral [30] where $p$ is constant. The solution of the hypercentral Schrödinger equation with Coulombic-like term plus a linear confining term potential cannot be obtained analytically [31] therefore, Giannini et al .used the dynamic symmetry O(7) of the hyperCoulomb problem to obtain the hyper Coulomb Hamiltonian and eigenfunctions analytically and they regarded the linear term as a perturbation. Combination of the color Coulomb plus linear confining term and the six-dimensional harmonic oscillator potential has interesting properties since it can be solved analytically, with a good correspondence to physical results. The first order correction $V^{(1)}(x)$ can be written as [30-33]

$$V^1(x) = -C_F C_A \frac{\alpha_s^2}{4x^2} \tag{2}$$

The parameters $C_F = \frac{2}{3}$ and $C_A = 3$ are the Casimir charges of the fundamental and adjoint representation. The hyper-coulomb strength $\tau = -\frac{2}{3}\alpha_s \cdot \frac{2}{3}$ is the color factor for the baryon. $\alpha_s$ is the strong running coupling constant, which is written as

$$\alpha_S = \frac{\alpha_S(\mu_0)}{1+\left(\frac{33-2n_f}{12\pi}\right)\alpha_S(\mu_0)\ln\left(\frac{m_1+m_2+m_3}{\mu_0}\right)} \qquad (3)$$

The spin-dependent part $V_{SD}(x)$ is given as

$$V_{SD}(x) = V_{SS}(x)(\vec{S}_\rho \cdot \vec{S}_\lambda) + V_{\gamma S}(x)(\vec{\gamma} \cdot \vec{S}) + V_T(x)\left[S^2 - \frac{3(\vec{S}\cdot\vec{x})(\vec{S}\cdot\vec{x})}{x^2}\right] \qquad (4)$$

The spin-dependent potential, $V_{SD}(x)$ contains three types of the interaction terms [34], such as the spin-spin term $V_{SS}(x)$, the spin-orbit term $V_{\gamma S}(x)$ and tensor term $V_T(x)$ described as [35]. Here $S = S_\rho + S_\lambda$ where $S_\rho$ and $S_\lambda$ are the spin vectors associated with the $\rho$ and $\lambda$ variables respectively. The coefficient of these spin-dependent terms of above equation can be written in terms of the vector, $V_V(x) = \frac{\tau}{x}$, and scalar, $V_S(x) = \beta x + px^2$ parts of the static potential as [25]

$$V_{\gamma S} = \frac{1}{2m_\rho m_\lambda x}\left(3\frac{dV_V}{dx} - \frac{dV_S}{dx}\right) \qquad (5)$$

$$V_T(x) = \frac{1}{6m_\rho m_\lambda}\left(\frac{3d^2V_V}{d^2x} - \frac{1}{x}\frac{dV_V}{dx}\right) \qquad (6)$$

$$V_{SS}(x) = \frac{1}{3m_\rho m_\lambda}\nabla^2 V_V \qquad (7)$$

In our model, the hypercentral interaction potential is assumed as follow [34]:

$$V(x) = V^{(0)}(x) + \left(\frac{1}{m_\rho} + \frac{1}{m_\lambda}\right)V^{(1)}(x) + V_{SD}(x) \qquad (8)$$

Where $V^{(0)}(x)$ is given by:

$$V^{(0)}(x) = V_{hyc}(x) + V_{con}(x) + V_{h.o.}(x) = \frac{\tau}{x} + \beta x + px^2 \qquad (9)$$

The baryons masses are determined by the sum of the model quark masses plus kinetic energy, potential energy and the spin-dependent interaction as $M_B = \sum m_i + \langle H \rangle$ [36]. First, we have solved the hyperradial Schrödinger equation exactly and find eigenvalue under the proposed potential by using the ansatz approach.

## 3. The Exact Analytical Solution of the Hyperradial Schrödinger Equation under the Hypercentral potential

The Hamiltonian of three bodies' baryonic system in the Hypercentral constituent quark model is expressed as [37]

$$H = \frac{P_\rho^2}{2m} + \frac{P_\lambda^2}{2m} + V(x) \tag{10}$$

and the hyperradial wave function $\psi_{v\gamma}(x)$ is determined by the hypercentral Schrödinger equation. The hyperradial Schrödinger equation corresponding to the above Hamiltonian can be written as [38]

$$\left(\frac{d^2}{dx^2} + \frac{5}{x}\frac{d}{dx} - \frac{\gamma(\gamma+4)}{x^2}\right)\psi_{v\gamma}(x) = -2m[E - V(x)]\psi_{v\gamma}(x) \tag{11}$$

Where $\gamma$ is the grand angular quantum number and given by $\gamma = 2n + l_\rho + l_\lambda$, $n = 0,1,...$; $l_\rho$ and $l_\lambda$ are the angular momenta associated with the $\vec{\rho}$ and $\vec{\lambda}$ variable and $V$ denotes the number of nodes of the space three-quark wave function [39]. In Eq. (11) $m$ is the reduced mass which is defined as $m = \frac{2m_\rho m_\lambda}{m_\rho + m_\lambda}$ [40]. By regarding $\psi_{v\gamma}(x) = x^{-\frac{5}{2}}\varphi_{v\gamma}$ [35, 41], Eq. (11) Reduce to the following form

$$\varphi_{v\gamma}''(x) + \left[\varepsilon - r_1 x^2 - r_2 x - \frac{r_3}{x} - \frac{r_4}{x^2} - \frac{r_5}{x^3} + \frac{r_6}{x^5} + r_7 - \frac{(2\gamma+3)(2\gamma+5)}{4x^2}\right]\varphi_{v\gamma}(x) = 0 \tag{12}$$

The hyperradial wave function $\varphi_{v\gamma}(x)$ is a solution of the reduced Schrödinger equation for each of the three identical particles with the mass m and interacting potential (8), where

$$\varepsilon = 2mE, \quad r_1 = 2mp, \quad r_2 = 2m\beta, \quad r_3 = 2m\tau,$$

$$r_4 = 2m\left(\frac{1}{m_\rho} + \frac{1}{m_\lambda}\right)\left(-C_f C_A \frac{\alpha_s^2}{4}\right), \quad r_5 = 2m\left[\frac{2\tau}{3m_\rho m_\lambda}(S_\rho \cdot S_\lambda) - \frac{3\tau}{2m_\rho m_\lambda}(\vec{\gamma}\cdot\vec{s}) + \frac{7\tau}{6m_\rho m_\lambda}s^2\right],$$

$$r_6 = 2m\frac{21\tau}{6m_\rho m_\lambda}(\vec{s}\cdot\vec{x})(\vec{s}\cdot\vec{x}), \quad r_7 = 2m\left(\frac{(\beta+2p)}{2m_\rho m_\lambda}(\vec{\gamma}\cdot\vec{s})\right), \tag{13}$$

We suppose the $\varphi_{v\gamma} = h(x)e^{g(x)}$ form for the wave function. Now we make use of the ansatz for the $h(x)$ and $g(x)$ [42-44]:

$$h(x) = \prod(x - a_i^v) \quad v = 1,2,...,$$

$$h(x) = 1 \quad v = 0$$

(14)

$$g(x) = a\ln x + qx^2 + cx + \frac{d}{x}$$

Where $a$, $q$, $c$ and $d$ are positive. From Eq. (14) we obtain

$$\varphi''(x) = \left[ g''(x) + g'^2(x) + \left(\frac{h''(x) + 2h'(x)g'(x)}{h(x)}\right) \right] \varphi(x) \tag{15}$$

Comparing Eqs. (12) and (15), it can be found that

$$\left[ r_1 x^2 + r_2 x + \frac{r_3}{x} + \frac{r_4}{x^2} + \frac{r_5}{x^3} - \frac{r_6}{x^5} - r_7 + \frac{(2\gamma+3)(2\gamma+5)}{4x^2} - \varepsilon \right] = \left[ g''(x) + g'^2(x) + \frac{h''(x) + 2h'(x)g'(x)}{h(x)} \right] \tag{16}$$

By substituting Eq. (14) into Eq. (16) we obtained the following equation

$$-\varepsilon + r_1 x^2 + r_2 x + \frac{r_3}{x} + \frac{r_4}{x^2} + \frac{r_5}{x^3} - \frac{r_6}{x^5} - r_7 + \frac{(2\gamma+3)(2\gamma+5)}{4x^2}$$

$$= 4q^2 x^2 + 4cqx + \frac{(2ac - 4dq)}{x} + \frac{(a^2 - a - 2cd)}{x^2} + \frac{2d(1-a)}{x^3} + \frac{d^2}{x^4} + (c^2 + 2q + 4ac) \tag{17}$$

By equating the corresponding powers of $x$ on both sides of Eq. (17), we can obtain

$$a = \frac{2\tau}{\beta}\sqrt{\frac{mp}{2}}, \quad c = \frac{m\beta}{2}\sqrt{\frac{2}{mp}}, \quad q = \sqrt{\frac{mp}{2}},$$

$$\varepsilon = -\left[ \frac{m\beta^2}{2p} + 2\sqrt{\frac{mp}{2}} + \frac{4mp\tau}{\beta} + 2m\left(\frac{(\beta+2p)}{2m_\rho m_\lambda}(\vec{\gamma}\cdot\vec{s})\right) \right] \tag{18}$$

Since $p = \frac{m\omega^2}{2}$, we have: $a = \frac{2m\omega}{2\beta}$, $c = \frac{\beta}{\omega}$, $q = \frac{m\omega}{2}$. The energy eigenvalues for the mode $\nu = 0$ and grand angular momentum $\gamma$ from Eqs. (13) and (18) are given as follows

$$E = -\left[ \frac{\beta^2}{2m\omega} + \frac{\omega}{2} + \frac{m\omega^2\tau}{\beta} + \left(\frac{(\beta+m\omega^2)}{2m_\rho m_\lambda}(\vec{\gamma}\cdot\vec{s})\right) \right] \tag{19}$$

At last for the best doubly heavy baryons masses ($\Xi_{ccd}$, $\Xi_{ccu}$, $\Xi_{bbd}$, $\Xi_{bbu}$, $\Xi_{bcd}$, $\Xi_{bcu}$) predictions, the values of $m_u$, $m_d$, $m_c$, $m_b$, $\alpha_S$, $\omega$ and $\beta$ (which are listed in Table 1) are selected using genetic algorithm. The cost function of a genetic algorithms the minimum difference between our calculated baryon mass and the reported baryons mass of other works.

**Table 1.** The Quark mass (in GeV) and the fitted values of the parameters used in our calculations.

| $m_b$ | $m_c$ | $m_d$ | $m_u$ | $\alpha_S$ | $C_F$ | $C_A$ | $\beta$ | $\omega$ |
|---|---|---|---|---|---|---|---|---|
| 4.750 | 1.348 | 0.35 | 0.34 | 0.340 | $\frac{2}{3}$ | 3 | 0.02 | 0.11 fm$^{-1}$ |

## 4. Results and Discussions: Mass Spectrum

The ground and excited states of doubly heavy $\Xi$ baryons are unclear to us experimentally (except $\Xi_{cc}^+$ and $\Xi_{cc}^{++}$). Hence, we have obtained the ground and excited state masses of $\Xi_{cc}^+$, $\Xi_{cc}^{++}$, $\Xi_{bb}^-$, $\Xi_{bb}^0$, $\Xi_{bc}^0$ and $\Xi_{bc}^+$ (see Tables 2, 3, 4, 5 and 6 respectively). These mass spectrum are estimated by using the hypercentral potential Eq. (8) in the hypercentral constituent quark model. We begin with the ground state $1S$, the masses are computed for both parities $J^P = \frac{1}{2}^+$ and $J^P = \frac{3}{2}^+$. Our predicted ground state masses of doubly heavy $\Xi$ baryons are compared with other predictions in Table 2.

**Table 2**. The outcomes ground state masses of $\Xi$ are listed with other theoretical predictions (in GeV). Standard devotion of the result is 0.350.

| Baryon | $\Xi_{ccd}$ / $\Xi_{ccu}$ | | $\Xi_{bbd}$ / $\Xi_{bbu}$ | | $\Xi_{bcd}$ / $\Xi_{bcu}$ | |
|---|---|---|---|---|---|---|
| $J^P$ | $\frac{1}{2}^+$ | $\frac{3}{2}^+$ | $\frac{1}{2}^+$ | $\frac{3}{2}^+$ | $\frac{1}{2}^+$ | $\frac{3}{2}^+$ |
| Our Calc | 3.522 / 3.515 | 3.696 / 3.689 | 9.716 / 9.711 | 9.894 / 9.889 | 6.628 / 6.622 | 6.688 / 6.682 |
| Ref.[1] | 3.520 / 3.511 | 3.695 / 3.687 | 10.317 / 10.312 | 10.340 / 10.335 | 6.920 / 6.914 | 6.986 / 6.980 |
| Ref.[45] | 3.519 | | | | | |
| Ref.[7] | 3.685 | 3.754 | 10.314 | | | |
| Ref.[12,13] | 3.520 | 3.695 | 10.199 | 10.316 | | |
| Ref.[5] | 3.610 | 3.694 | | | | |
| Ref.[14] | 3.610 | 3.692 | 10.143 | 10.178 | 6.943 | 6.985 |
| Ref.[46] | 3.561 | 3.642 | | | | |
| Ref.[17] | 3.720 | | 9.960 | | 6.720 | |
| Ref.[18] | 3.687 | 3.752 | 10.322 | 10.352 | 7.014 | 7.064 |
| Ref.[47] | 3.676 | 3.753 | 10.340 | 10.367 | 7.011 | 7.074 |
| Ref.[48] | 3.547 | 3.719 | 10.185 | 10.216 | 6.904 | 6.936 |
| Ref.[19] | 3.579 | 3.656 | 10.189 | 10.218 | | |
| Ref.[20] | 3.620 | 3.727 | 10.202 | 10.237 | 6.933 | 6.980 |
| Ref.[21] | 3.478 | 3.610 | 10.093 | 10.133 | 6.820 | 6.900 |
| Ref.[49] | 3.627 | 3.690 | 10.162 | 10.184 | 6.914 | |
| Ref.[50] | 3.519 | 3.620 | 9.800 | 9.980 | 6.650 | 6.690 |
| Ref.[22] | 3.612 | 3.706 | 10.197 | 10.136 | 6.919 | 6.986 |
| Ref.[51] | 3.510 | 3.548 | 10.130 | 10.144 | 6.792 | 6.827 |
| Ref.[52] | 3.570 | 3.610 | 10.170 | 10.220 | | |

**Table 3.** The masses of radial excited states for doubly heavy $\Xi$ baryons (in GeV). Standard devotions of the result are 0.435 and 0.434.

| Baryon | State | $J^P$ | Our Calc | Our Calc | [1] | [1] | [7] | [47] | [48] | [19] | [20] | [18] |
|---|---|---|---|---|---|---|---|---|---|---|---|---|
| $\Xi_{ccd}$ and $\Xi_{ccu}$ | 2S | $\frac{1}{2}^+$ | 3.905 | 3.901 | 3.925 | 3.920 | 4.079 | 4.029 | 4.183 | 3.976 | 3.910 | 4.030 |
| | 3S | | 4.185 | 4.118 | 4.233 | 4.159 | 4.206 | | 4.640 | | 4.154 | |
| | 4S | | 4.430 | 4.429 | 4.502 | 4.501 | | | | | | |
| | 5S | | 4.653 | 4.653 | 4.748 | 4.748 | | | | | | |
| | 2S | $\frac{3}{2}^+$ | 3.962 | 3.958 | 3.988 | 3.983 | 4.114 | 4.042 | 4.282 | 4.025 | 4.027 | 4.078 |
| | 3S | | 4.213 | 4.211 | 4.264 | 4.261 | 4.131 | | 4.719 | | | |
| | 4S | | 4.446 | 4.445 | 4.520 | 4.519 | | | | | | |
| | 5S | | 4.663 | 4.663 | 4.759 | 4.759 | | | | | | |
| $\Xi_{bbd}$ and $\Xi_{bbu}$ | 2S | $\frac{1}{2}^+$ | 9.984 | 9.981 | 10.612 | 10.609 | 10.571 | 10.576 | 10.751 | 10.482 | 10.441 | 10.551 |
| | 3S | | 10.211 | 10.211 | 10.862 | 10.862 | 10.612 | | 11.170 | | 10.630 | |
| | 4S | | 10.417 | 10.418 | 11.088 | 11.090 | | | | | 10.812 | |
| | 5S | | 10.606 | 10.610 | 11.297 | 11.301 | | | | | | |
| | 2S | $\frac{3}{2}^+$ | 9.990 | 9.988 | 10.619 | 10.617 | 10.592 | 10.578 | 10.770 | 10.501 | 10.482 | 10.574 |
| | 3S | | 10.205 | 10.233 | 10.855 | 10.866 | 10.593 | | 11.184 | | 10.673 | |
| | 4S | | 10.418 | 10.420 | 11.090 | 11.092 | | | | | 10.856 | |
| | 5S | | 10.607 | 10.611 | 11.298 | 11.302 | | | | | | |
| $\Xi_{bcd}$ and $\Xi_{bcu}$ | 2S | $\frac{1}{2}^+$ | 6.922 | 6.919 | 7.244 | 7.240 | | | 7.478 | | | 7.321 |
| | 3S | | 7.163 | 7.161 | 7.509 | 7.507 | | | 7.904 | | | |
| | 4S | | 7.379 | 7.377 | 7.746 | 7.744 | | | | | | |
| | 5S | | 7.576 | 7.581 | 7.963 | 7.964 | | | | | | |
| | 2S | $\frac{3}{2}^+$ | 6.943 | 6.939 | 7.267 | 7.263 | | | 7.495 | | | 7.353 |
| | 3S | | 7.174 | 7.171 | 7.521 | 7.518 | | | 7.917 | | | |
| | 4S | | 7.384 | 7.384 | 7.752 | 7.752 | | | | | | |
| | 5S | | 7.580 | 7.581 | 7.968 | 7.969 | | | | | | |

We can observe that in the case of $\Xi_{cc}$ baryon, for 2S states $J^P = \frac{1}{2}^+$ and $J^P = \frac{3}{2}^+$ our predictions are close to Ref. [43] and Ref. [1], respectively. Our outcomes for 3S state $J^P = \frac{1}{2}^+$ of $\Xi_{cc}$ baryon shows 21 MeV (with [7]) and $J^P = \frac{3}{2}^+$ shows 51 MeV (with [1]) difference. Analyzing the 2S and 3S states masses for $\Xi_{bb}$ and $\Xi_{bc}$ baryons (with both parities) show that our masses have a difference in the range of $\approx 0.5$ GeV with Refs. [1, 7, 39-43].

To calculate the orbital excited state masses (1P–5P, 1D– 4D, 1F–2F) we have considered all possible isospin splitting and all combinations of total spin S and total angular momentum J. Our outcomes and the comparison of masses with other approaches are also tabulated in Tables 4, 5 and 6.

**Table 4.** The masses of orbital excited for $\Xi_{cc}$ baryon (in GeV). Standard devotions of the result are 0.072 and 0.068.

| State | Our cal $\Xi_{cc}^+$ | Our Cal $\Xi_{cc}^{++}$ | [1] $\Xi_{cc}^+$ | [1] $\Xi_{cc}^{++}$ | [7] | [47] | [19] | [20] | [12] | [21] | [18] | [5] |
|---|---|---|---|---|---|---|---|---|---|---|---|---|
| $(1^2P_{1/2})$ | 3.851 | 3.847 | 3.865 | 3.861 | 3.947 | 3.910 | 3.880 | 3.838 | | | 4.073 | 3.892 |
| $(1^2P_{3/2})$ | 3.834 | 3.830 | 3.847 | 3.842 | 3.949 | 3.921 | | 3.959 | 3.786 | 3.834 | 4.079 | 3.989 |
| $(1^4P_{1/2})$ | 3.860 | 3.856 | 3.875 | 3.871 | | | | | | | | |
| $(1^4P_{3/2})$ | 3.842 | 3.838 | 3.856 | 3.851 | | | | | | | | |
| $(1^4P_{5/2})$ | 3.873 | 3.872 | 3.890 | 3.888 | 4.163 | 4.092 | | 4.155 | 3.949 | 4.047 | 4.089 | |
| $(2^2P_{1/2})$ | 4.120 | 4.101 | 4.161 | 4.140 | 4.135 | 4.074 | 4.018 | 4.085 | | | | |
| $(2^2P_{3/2})$ | 4.104 | 4.101 | 4.144 | 4.140 | 4.137 | 4.078 | 4.197 | | | | | |
| $(2^4P_{1/2})$ | 4.127 | 4.125 | 4.169 | 4.167 | | | | | | | | |
| $(2^4P_{3/2})$ | 4.111 | 4.109 | 4.152 | 4.149 | | | | | | | | |
| $(2^4P_{5/2})$ | 4.140 | 4.138 | 4.183 | 4.181 | 4.488 | | | | | | | |
| $(3^2P_{1/2})$ | 4.361 | 4.345 | 4.426 | 4.409 | 4.149 | | | | | | | |
| $(3^2P_{3/2})$ | 4.347 | 4.345 | 4.411 | 4.409 | 4.159 | | | | | | | |
| $(3^4P_{1/2})$ | 4.367 | 4.366 | 4.433 | 4.432 | | | | | | | | |
| $(3^4P_{3/2})$ | 4.354 | 4.352 | 4.419 | 4.417 | | | | | | | | |
| $(3^4P_{5/2})$ | 4.336 | 4.333 | 4.399 | 4.396 | 4.534 | | | | | | | |
| $(4^2P_{1/2})$ | 4.583 | 4.583 | 4.671 | 4.671 | | | | | | | | |
| $(4^2P_{3/2})$ | 4.571 | 4.571 | 4.658 | 4.657 | | | | | | | | |
| $(4^4P_{1/2})$ | 4.590 | 4.590 | 4.678 | 4.678 | | | | | | | | |
| $(4^4P_{3/2})$ | 4.577 | 4.577 | 4.664 | 4.664 | | | | | | | | |
| $(4^4P_{5/2})$ | 4.561 | 4.561 | 4.646 | 4.646 | | | | | | | | |
| $(5^2P_{1/2})$ | 4.792 | 4.793 | 4.901 | 4.902 | | | | | | | | |
| $(5^2P_{3/2})$ | 4.781 | 4.781 | 4.889 | 4.889 | | | | | | | | |
| $(5^4P_{1/2})$ | 4.799 | 4.800 | 4.908 | 4.909 | | | | | | | | |
| $(5^4P_{3/2})$ | 4.705 | 4.788 | 4.895 | 4.896 | | | | | | | | |
| $(5^4P_{5/2})$ | 4.771 | 4.772 | 4.878 | 4.879 | | | | | | | | |
| $(1^4D_{1/2})$ | 4.043 | 4.038 | 4.077 | 4.071 | | | | | | | | |
| $(1^2D_{3/2})$ | 4.018 | 4.013 | 4.049 | 4.044 | | | | | | | | |
| $(1^4D_{3/2})$ | 4.026 | 4.022 | 4.058 | 4.053 | | | | | | | | |
| $(1^2D_{5/2})$ | 3.995 | 3.991 | 4.024 | 4.019 | 4.043 | 4.115 | 4.047 | | 4.391 | 4.034 | 4.050 | 4.388 |
| $(1^4D_{5/2})$ | 4.003 | 4.000 | 4.033 | 4.029 | 4.027 | 4.052 | | 4.187 | 4.089 | 4.393 | | |
| $(1^4D_{7/2})$ | 3.975 | 3.972 | 4.002 | 3.998 | 4.097 | | | | | | | |
| $(2^4D_{1/2})$ | 4.287 | 4.284 | 4.345 | 4.342 | | | | | | | | |
| $(2^2D_{3/2})$ | 4.265 | 4.262 | 4.321 | 4.318 | | | | | | | | |
| $(2^4D_{3/2})$ | 4.272 | 4.270 | 4.329 | 4.326 | | | | | | | | |
| $(2^2D_{5/2})$ | 4.245 | 4.243 | 4.299 | 4.297 | 4.164 | 4.091 | | | | | | |
| $(2^4D_{5/2})$ | 4.252 | 4.251 | 4.307 | 4.305 | | | | | | | | |
| $(2^4D_{7/2})$ | 4.228 | 4.226 | 4.280 | 4.278 | 4.394 | | | | | | | |

**Table 4.** Continue.

| State | Our Cal $\Xi_{cc}^+$ | Our Cal $\Xi_{cc}^{++}$ | [1] $\Xi_{cc}^+$ | [1] $\Xi_{cc}^{++}$ | [7] | [40] | [42] | [43] | [35] | [44] | [39] | [5] |
|---|---|---|---|---|---|---|---|---|---|---|---|---|
| $(3^4 D_{1/2})$ | 4.511 | 4.511 | 4.592 | 4.592 | 4.511 | 4.511 | 4.592 | 4.592 | 4.511 | 4.511 | 4.592 | 4.592 |
| $(3^2 D_{3/2})$ | 4.492 | 4.491 | 4.571 | 4.570 | 4.492 | 4.491 | 4.571 | 4.570 | 4.492 | 4.491 | 4.571 | 4.570 |
| $(3^4 D_{3/2})$ | 4.499 | 4.499 | 4.578 | 4.578 | | | | | | | | |
| $(3^2 D_{5/2})$ | 4.475 | 4.474 | 4.552 | 4.551 | 4.348 | | | | | | | |
| $(3^4 D_{5/2})$ | 4.481 | 4.481 | 4.559 | 4.558 | | | | | | | | |
| $(3^4 D_{7/2})$ | 4.460 | 4.459 | 4.535 | 4.534 | | | | | | | | |
| $(4^4 D_{1/2})$ | 4.723 | 4.724 | 4.825 | 4.826 | | | | | | | | |
| $(4^2 D_{3/2})$ | 4.706 | 4.706 | 4.806 | 4.806 | | | | | | | | |
| $(4^4 D_{3/2})$ | 4.711 | 4.712 | 4.812 | 4.813 | | | | | | | | |
| $(4^2 D_{5/2})$ | 4.690 | 4.690 | 4.788 | 4.788 | | | | | | | | |
| $(4^4 D_{5/2})$ | 4.696 | 4.696 | 4.795 | 4.795 | | | | | | | | |
| $(4^4 D_{7/2})$ | 4.675 | 4.675 | 4.772 | 4.772 | | | | | | | | |
| $(1^4 F_{3/2})$ | 4.198 | 4.193 | 4.247 | 4.242 | | | | | | | | |
| $(1^2 F_{5/2})$ | 4.169 | 4.164 | 4.215 | 4.210 | | | | | | | | |
| $(1^4 F_{5/2})$ | 4.142 | 4.172 | 4.186 | 4.219 | | | | | | | | |
| $(1^4 F_{7/2})$ | 4.150 | 4.147 | 4.194 | 4.191 | | | | | | | | |
| $(1^2 F_{7/2})$ | 4.178 | 4.139 | 4.225 | 4.182 | | | | | | 4.267 | | |
| $(1^4 F_{9/2})$ | 4.118 | 4.115 | 4.159 | 4.156 | | | | | | 4.413 | | |
| $(2^4 F_{3/2})$ | 4.422 | 4.425 | 4.494 | 4.497 | | | | | | | | |
| $(2^2 F_{5/2})$ | 4.399 | 4.399 | 4.468 | 4.468 | | | | | | | | |
| $(2^4 F_{5/2})$ | 4.405 | 4.406 | 4.475 | 4.476 | | | | | | | | |
| $(2^4 F_{7/2})$ | 4.378 | 4.382 | 4.445 | 4.450 | | | | | | | | |
| $(2^2 F_{7/2})$ | 4.384 | 4.376 | 4.452 | 4.443 | | | | | | | | |
| $(2^4 F_{9/2})$ | 4.359 | 4.355 | 4.424 | 4.420 | | | | | | | | |

Our obtained orbital excited masses for $\Xi_{cc}$, 1P state $J^P = \frac{1}{2}^-$ shows a difference of 14 MeV (with [1]), 29 MeV (with [42]), 13 MeV (with [43]) and 41 MeV (with [5]), while 1P state $J^P = \frac{3}{2}^-$ shows 14MeV (with [1]), 48 MeV (with [35]) and 0 MeV (with [44]). Our 2P state $J^P = \frac{1}{2}^-$ shows a difference of 15 MeV (with [7]), 35 MeV (with [43]) and 41 MeV (with [1]), while 2P state $J^P = \frac{3}{2}^-$ shows 26MeV (with [40]), 33 MeV (with [7]) and 40 MeV (with [1]). Results for 3P states $J^P = \frac{1}{2}^-$ and $J^P = \frac{3}{2}^-$ show a difference in the range of $\approx$ 60MeV with Ref. [1]. We can easily observe that our calculated masses for 4P-5P, 1D-3D and 1F-2F are matched with Ref. [1]. Our outcome for 3D state $J^P = \frac{3}{2}^+$ is quite equal to the predictions of Refs. [7, 40, 35, 44]. For the ground and excited states of doubly heavy baryons ($\Xi_{cc}^+$), the minimum and maximum percentage of relative error values are 0% and 3.53% between our calculations and the masses reported by Shah et al. [1].

**Table 5.** The masses of orbital excited states for $\Xi_{bb}$ baryon (in GeV).

| State | Our cal $\Xi_{bb}^-$ | Our Cal $\Xi_{bb}^0$ | [1] $\Xi_{bb}^-$ | [1] $\Xi_{bb}^0$ | [7] | [47] | [19] | [20] | [12] | [18] | Others |
|---|---|---|---|---|---|---|---|---|---|---|---|
| $(1^2 P_{1/2})$ | 9.895 | 9.892 | 10.514 | 10.511 | 10.476 | 10.493 | 10.406 | 10.368 |  | 10.691 |  |
| $(1^2 P_{3/2})$ | 9.890 | 9.887 | 10.509 | 10.506 | 10.476 | 10.495 |  | 10.408 | 10.474 | 10.692 | 10.390 [52] |
| $(1^4 P_{1/2})$ | 9.897 | 9.895 | 10.517 | 10.514 |  |  |  |  |  |  |  |
| $(1^4 P_{3/2})$ | 9.893 | 9.890 | 10.512 | 10.509 |  |  |  |  |  |  | 10.430 [17] |
| $(1^4 P_{5/2})$ | 9.901 | 9.898 | 10.521 | 10.518 | 10.759 |  |  |  | 10.588 | 10.695 |  |
| $(2^2 P_{1/2})$ | 10.127 | 10.127 | 10.77 | 10.77 | 10.703 | 10.710 | 10612 | 10.563 |  |  |  |
| $(2^2 P_{3/2})$ | 10.124 | 10.120 | 10.766 | 10.762 | 10.704 | 10.713 |  | 10.607 |  |  |  |
| $(2^4 P_{1/2})$ | 10.129 | 10.129 | 10.772 | 10.772 |  |  |  |  |  |  |  |
| $(2^4 P_{3/2})$ | 10.126 | 10.125 | 10.768 | 10.767 |  |  |  |  |  |  |  |
| $(2^4 P_{5/2})$ | 10.121 | 10.133 | 10.763 | 10.776 | 10.973 | 10.713 |  |  |  |  |  |
| $(3^2 P_{1/2})$ | 10.337 | 10.338 | 11.001 | 11.002 | 10.740 |  |  | 10.744 |  |  |  |
| $(3^2 P_{3/2})$ | 10.334 | 10.335 | 10.997 | 10.998 | 10.742 |  |  | 10.788 |  |  |  |
| $(3^4 P_{1/2})$ | 10.339 | 10.340 | 11.003 | 11.004 |  |  |  |  |  |  |  |
| $(3^4 P_{3/2})$ | 10.336 | 10.337 | 10.999 | 11.000 |  |  |  |  |  |  |  |
| $(3^4 P_{5/2})$ | 10.331 | 10.343 | 10.994 | 11.007 | 11.004 |  |  |  |  |  |  |
| $(4^2 P_{1/2})$ | 10.531 | 10.534 | 11.214 | 11.217 |  |  |  | 10.900 |  |  |  |
| $(4^2 P_{3/2})$ | 10.527 | 10.530 | 11.21 | 11.213 |  |  |  |  |  |  |  |
| $(4^4 P_{1/2})$ | 10.533 | 10.536 | 11.216 | 11.219 |  |  |  |  |  |  |  |
| $(4^4 P_{3/2})$ | 10.529 | 10.532 | 11.212 | 11.215 |  |  |  |  |  |  |  |
| $(4^4 P_{5/2})$ | 10.526 | 10.538 | 11.208 | 11.222 |  |  |  |  |  |  |  |
| $(5^2 P_{1/2})$ | 10.712 | 10.716 | 11.413 | 11.418 |  |  |  |  |  |  |  |
| $(5^2 P_{3/2})$ | 10.709 | 10.714 | 11.41 | 11.415 |  |  |  |  |  |  |  |
| $(5^4 P_{1/2})$ | 10.714 | 10.718 | 11.415 | 11.420 |  |  |  |  |  |  |  |
| $(5^4 P_{3/2})$ | 10.711 | 10.716 | 11.412 | 11.417 |  |  |  |  |  |  |  |
| $(5^4 P_{5/2})$ | 10.706 | 10.721 | 11.407 | 11.423 |  |  |  |  |  |  |  |
| $(1^4 D_{1/2})$ | 10.043 | 10.041 | 10.677 | 10.675 |  |  |  |  |  |  |  |
| $(1^2 D_{3/2})$ | 10.037 | 10.035 | 10.670 | 10.668 |  |  |  |  |  |  |  |
| $(1^4 D_{3/2})$ | 10.038 | 10.037 | 10.672 | 10.670 |  |  |  |  |  | 11.011 |  |
| $(1^2 D_{5/2})$ | 10.030 | 10.028 | 10.663 | 10.661 | 10.592 | 10.676 |  |  | 10.742 | 11.002 |  |
| $(1^4 D_{5/2})$ | 10.033 | 10.031 | 10.666 | 10.664 |  |  |  |  |  |  |  |
| $(1^4 D_{7/2})$ | 10.026 | 10.024 | 10.658 | 10.656 |  | 10.608 |  |  | 10.853 | 11.011 |  |
| $(2^4 D_{1/2})$ | 10.257 | 10.257 | 10.913 | 10.913 |  |  |  |  |  |  |  |
| $(2^2 D_{3/2})$ | 10.252 | 10.252 | 10.907 | 10.907 |  |  |  |  |  |  |  |
| $(2^4 D_{3/2})$ | 10.254 | 10.254 | 10.909 | 10.909 |  |  |  |  |  |  |  |
| $(2^2 D_{5/2})$ | 10.247 | 10.247 | 10.901 | 10.901 |  | 10.712 |  |  |  |  |  |
| $(2^4 D_{5/2})$ | 10.248 | 10.248 | 10.903 | 10.903 | 10.613 |  |  |  |  |  |  |
| $(2^4 D_{7/2})$ | 10.242 | 10.242 | 10.896 | 10.896 |  | 11.057 |  |  |  |  |  |

**Table 5**. Continue.

| State | Our cal $\Xi_{bb}^-$ | Our Cal $\Xi_{bb}^0$ | [1] $\Xi_{bb}^-$ | [1] $\Xi_{bb}^0$ | [7] | [47] | [19] | [20] | [12] | [18] | Others |
|---|---|---|---|---|---|---|---|---|---|---|---|
| $(3^4 D_{1/2})$ | 10.455 | 10.457 | 11.13 | 11.133 | | | 4.592 | 4.592 | | | |
| $(3^2 D_{3/2})$ | 10.450 | 10.452 | 11.125 | 11.127 | | | 4.571 | 4.570 | | | |
| $(3^4 D_{3/2})$ | 10.451 | 10.454 | 11.126 | 11.129 | | | | | | | |
| $(3^2 D_{5/2})$ | 10.446 | 10.447 | 11.120 | 11.122 | | | | | | | |
| $(3^4 D_{5/2})$ | 10.447 | 10.449 | 11.122 | 11.124 | 10.809 | | | | | | |
| $(3^4 D_{7/2})$ | 10.442 | 10.444 | 11.116 | 11.118 | | | | | | | |
| $(4^4 D_{1/2})$ | 10.639 | 10.643 | 11.333 | 11.337 | | | | | | | |
| $(4^2 D_{3/2})$ | 10.635 | 10.638 | 11.328 | 11.332 | | | | | | | |
| $(4^4 D_{3/2})$ | 10.636 | 10.640 | 11.330 | 11.334 | | | | | | | |
| $(4^2 D_{5/2})$ | 10.631 | 10.635 | 11.324 | 11.328 | | | | | | | |
| $(4^4 D_{5/2})$ | 10.632 | 10.636 | 11.325 | 11.33 | | | | | | | |
| $(4^4 D_{7/2})$ | 10.627 | 10.631 | 11.320 | 11.324 | | | | | | | |
| $(1^4 F_{3/2})$ | 10.173 | 10.172 | 10.82 | 10.819 | | | | | | | |
| $(1^2 F_{5/2})$ | 10.166 | 10.165 | 10.812 | 10.811 | | | | | | | |
| $(1^4 F_{5/2})$ | 10.158 | 10.167 | 10.804 | 10.813 | | | | | | | |
| $(1^4 F_{7/2})$ | 10.167 | 10.160 | 10.814 | 10.806 | | | | | | | |
| $(1^2 F_{7/2})$ | 10.160 | 10.157 | 10.806 | 10.803 | | | | | 11.004 | | |
| $(1^4 F_{9/2})$ | 10.152 | 10.152 | 10.797 | 10.797 | | | | | 11.112 | | |
| $(2^4 F_{3/2})$ | 10.357 | 10.376 | 11.022 | 11.043 | | | | | | | |
| $(2^2 F_{5/2})$ | 10.368 | 10.369 | 11.035 | 11.036 | | | | | | | |
| $(2^4 F_{5/2})$ | 10.369 | 10.371 | 11.036 | 11.038 | | | | | | | |
| $(2^4 F_{7/2})$ | 10.362 | 10.365 | 11.028 | 11.031 | | | | | | | |
| $(2^2 F_{7/2})$ | 10.364 | 10.363 | 11.030 | 11.029 | | | | | | | |
| $(2^4 F_{9/2})$ | 10.357 | 10.357 | 11.022 | 11.023 | | | | | | | |

For $\Xi_{bb}$ and $\Xi_{bc}$ baryons, the mass difference from our calculations and other references is large.

Comparing our findings with the masses reported by Shah et al. [1], the minimum and maximum percentage of relative error values are 1.2% (0.8%) and 10.317% (6.92%) for the ground and excited states of doubly heavy baryons $\Xi_{bb}$ and $\Xi_{bc}$, respectively.

**Table 6.** The masses of orbital excited states for $\Xi_{bc}$ baryon (in GeV).

| State | Our cal $\Xi_{bc}^0$ | Our Cal $\Xi_{bc}^+$ | [1] $\Xi_{bc}^0$ | [1] $\Xi_{bc}^+$ | [18] |
|---|---|---|---|---|---|
| $(1^2P_{1/2})$ | 6.846 | 6.842 | 7.16 | 7.156 | 7.390 |
| $(1^2P_{3/2})$ | 6.836 | 6.831 | 7.149 | 7.144 | 7.394 |
| $(1^4P_{1/2})$ | 6.851 | 6.847 | 7.166 | 7.161 | 7.399 |
| $(1^4P_{3/2})$ | 6.841 | 6.837 | 7.155 | 7.15 | |
| $(1^4P_{5/2})$ | 6.859 | 6.856 | 7.175 | 7.171 | |
| $(2^2P_{1/2})$ | 7.087 | 7.084 | 7.425 | 7.422 | |
| $(2^2P_{3/2})$ | 7.078 | 7.075 | 7.415 | 7.412 | |
| $(2^4P_{1/2})$ | 7.091 | 7.088 | 7.43 | 7.426 | |
| $(2^4P_{3/2})$ | 7.082 | 7.079 | 7.42 | 7.417 | |
| $(2^4P_{5/2})$ | 7.071 | 7.095 | 7.408 | 7.434 | |
| $(3^2P_{1/2})$ | 7.304 | 7.302 | 7.664 | 7.662 | |
| $(3^2P_{3/2})$ | 7.296 | 7.295 | 7.655 | 7.654 | |
| $(3^4P_{1/2})$ | 7.308 | 7.306 | 7.668 | 7.666 | |
| $(3^4P_{3/2})$ | 7.299 | 7.299 | 7.659 | 7.658 | |
| $(3^4P_{5/2})$ | 7.289 | 7.312 | 7.648 | 7.673 | |
| $(4^2P_{1/2})$ | 7.504 | 7.623 | 7.884 | 8.015 | |
| $(4^2P_{3/2})$ | 7.497 | 7.498 | 7.876 | 7.877 | |
| $(4^4P_{1/2})$ | 7.508 | 7.508 | 7.888 | 7.888 | |
| $(4^4P_{3/2})$ | 7.500 | 7.500 | 7.88 | 7.88 | |
| $(4^4P_{5/2})$ | 7.491 | 7.514 | 7.87 | 7.895 | |
| $(5^2P_{1/2})$ | 7.692 | 7.693 | 8.091 | 8.092 | |
| $(5^2P_{3/2})$ | 7.686 | 7.687 | 8.084 | 8.085 | |
| $(5^4P_{1/2})$ | 7.695 | 7.697 | 8.094 | 8.096 | |
| $(5^4P_{3/2})$ | 7.689 | 7.689 | 8.087 | 8.088 | |
| $(5^4P_{5/2})$ | 7.680 | 7.681 | 8.078 | 8.079 | |
| $(1^4D_{1/2})$ | 7.006 | 7.004 | 7.336 | 7.334 | |
| $(1^2D_{3/2})$ | 6.992 | 6.989 | 7.321 | 7.318 | |
| $(1^4D_{3/2})$ | 6.997 | 6.980 | 7.326 | 7.308 | 7.324 |
| $(1^2D_{5/2})$ | 6.980 | 6.977 | 7.308 | 7.304 | |
| $(1^4D_{5/2})$ | 6.985 | 6.969 | 7.313 | 7.295 | 7.309 |
| $(1^4D_{7/2})$ | 6.969 | 6.953 | 7.296 | 7.278 | 7.292 |
| $(2^4D_{1/2})$ | 7.087 | 7.227 | 7.425 | 7.579 | 7.579 |
| $(2^2D_{3/2})$ | 7.216 | 7.214 | 7.567 | 7.565 | |
| $(2^4D_{3/2})$ | 7.219 | 7.219 | 7.571 | 7.57 | |
| $(2^2D_{5/2})$ | 7.205 | 7.203 | 7.555 | 7.553 | 7.538 |
| $(2^4D_{5/2})$ | 7.209 | 7.208 | 7.559 | 7.558 | |
| $(2^4D_{7/2})$ | 7.196 | 7.195 | 7.545 | 7.544 | |

**Table 6**. Continue.

| State | Our cal $\Xi_{bc}^0$ | Our Cal $\Xi_{bc}^+$ | [1] $\Xi_{bc}^0$ | [1] $\Xi_{bc}^+$ | [18] |
|---|---|---|---|---|---|
| $(3^4 D_{1/2})$ | 7.431 | 7.431 | 7.804 | 7.804 | |
| $(3^2 D_{3/2})$ | 7.420 | 7.420 | 7.792 | 7.792 | |
| $(3^4 D_{3/2})$ | 7.411 | 7.424 | 7.782 | 7.796 | |
| $(3^2 D_{5/2})$ | 7.415 | 7.410 | 7.786 | 7.781 | |
| $(3^4 D_{5/2})$ | 7.402 | 7.414 | 7.772 | 7.785 | |
| $(3^4 D_{7/2})$ | 7.402 | 7.402 | 7.772 | 7.772 | |
| $(4^4 D_{1/2})$ | 7.429 | 7.504 | 7.801 | 7.884 | 7.797 |
| $(4^2 D_{3/2})$ | 7.611 | 7.613 | 8.002 | 8.004 | |
| $(4^4 D_{3/2})$ | 7.615 | 7.617 | 8.006 | 8.008 | |
| $(4^2 D_{5/2})$ | 7.603 | 7.604 | 7.993 | 7.994 | |
| $(4^4 D_{5/2})$ | 7.606 | 7.608 | 7.996 | 7.998 | |
| $(4^4 D_{7/2})$ | 7.596 | 7.597 | 7.985 | 7.986 | |
| $(1^4 F_{3/2})$ | 7.143 | 7.141 | 7.487 | 7.485 | |
| $(1^2 F_{5/2})$ | 7.127 | 7.125 | 7.469 | 7.467 | |
| $(1^4 F_{5/2})$ | 7.131 | 7.129 | 7.474 | 7.472 | |
| $(1^4 F_{7/2})$ | 7.117 | 7.114 | 7.458 | 7.455 | |
| $(1^2 F_{7/2})$ | 7.112 | 7.109 | 7.453 | 7.45 | |
| $(1^4 F_{9/2})$ | 7.099 | 7.097 | 7.439 | 7.436 | |
| $(2^4 F_{3/2})$ | 7.350 | 7.350 | 7.715 | 7.715 | |
| $(2^2 F_{5/2})$ | 7.337 | 7.336 | 7.7 | 7.699 | |
| $(2^4 F_{5/2})$ | 7.340 | 7.339 | 7.704 | 7.703 | |
| $(2^4 F_{7/2})$ | 7.328 | 7.327 | 7.69 | 7.689 | |
| $(2^2 F_{7/2})$ | 7.324 | 7.323 | 7.686 | 7.685 | |
| $(2^4 F_{9/2})$ | 7.313 | 7.311 | 7.674 | 7.672 | |

## 6. Conclusion

In this study, we have computed the mass spectra of ground and excited states for doubly heavy $\Xi$ baryons by using a hypercentral constituent quark model. For this goal we have analytically solved the hyperradial Schrödinger equation for three identical interacting particles under the effective hypercentral potential by using the ansatz method. Our proposed potential is regarded as a combination of the Coulombic-like term plus a linear confining term and the harmonic oscillator potential. We also added the first order correction and the spin-dependent part to the potential. In our calculations, the $u$ and $d$ quarks have 10 MeV difference mass, so there is a very small mass difference between $\Xi_{ccd}$ and $\Xi_{ccu}$, $\Xi_{bbd}$ and $\Xi_{bbu}$, $\Xi_{bcd}$ and $\Xi_{bcu}$. Our model has succeeded to assign the $J^P$ values to the exited states of doubly heavy baryons ($\Xi_{ccd}$, $\Xi_{ccu}$, $\Xi_{bbd}$, $\Xi_{bbu}$, $\Xi_{bcd}$ and $\Xi_{bcu}$). Comparison of the results with other predictions revealed that they are in agreement and our proposed model can be useful to investigate the doubly heavy baryons states masses. For example, for the ground, radial and orbital excited states masses of doubly heavy $\Xi$ baryons the minimum and the maximum percentage of relative error values are 0% and 6% between our calculations and the masses reported by Shah et al. [1].

**Acknowledgment**

The authors thank the referees for a thorough reading of our manuscript and for constructive suggestion.